\documentstyle[aps,preprint]{revtex}

\begin{document}
\draft \preprint{quant-ph/9902047} \tightenlines
\title{One-Parameter Gaussian State of Anharmonic Oscillator:
Nonlinear Realization of Bogoliubov Transformation}

\author{Jung Kon Kim, Sang Pyo Kim\footnote{sangkim@knusun1.kunsan.ac.kr},
Kwang-Gi Park}

\address{Department of Physics,
Kunsan National University,
Kunsan 573-701, Korea}
\date{\today}

\maketitle
\begin{abstract}
We find a one-parameter Gaussian state for an anharmonic
oscillator with quadratic and quartic terms,
which depends on the energy expectation value.
For the weak coupling constant, the Gaussian state
is a squeezed state of the vacuum state.
However, for the strong coupling constant, the Gaussian state represents
a different kind of condensation of bosonic particles through
a nonlinear Bogoliubov transformation of the vacuum state.
\end{abstract}

Harmonic oscillator has been used to describe a non-interacting ideal
bosonic gas. One of physically interesting quantum states is
the Gaussian state. Recently we have found a one-parameter squeezed
Gaussian state for a time-independent harmonic oscillator
whose squeezing parameter is the energy expectation value \cite{kim1}.
It was also shown there that even for a time-dependent
harmonic oscillator this one-parameter Gaussian state is still
the squeezed state of the vacuum state with the minimum
uncertainty. This squeezed Gaussian state forms a condensation (squeezed vacuum)
of bosonic particles, and represents a higher energy state.

On the other hand, anharmonic oscillator, as a $(0+1)$-dimensional toy model for a quantum
field, describes a self-interacting bosonic gas. Therefore it would be physically
interesting to have an analogous Gaussian state for the anharmonic oscillator.
In this paper we find a one-parameter Gaussian state for the anharmonic oscillator
whose parameter depends on the energy expectation value, and study the relation between
the Gaussian state and the (approximate) vacuum state. It is found
that for the weak coupling constant the one-parameter Gaussian state
is a squeezed vacuum of the vacuum state, but for the strong coupling
constant it can not be obtained from the vacuum state through a {\it linear}
Bogoliubov transformation. That is, the Gaussian state for the strong
coupling constant does not represent merely the squeezed vacuum, rather
it exhibits a different kind of condensation of bosonic particles
through a {\it nonlinear} Bogoliubov transformation.

We now wish to find the Gaussian state for an
anharmonic oscillator with quadratic and quartic potential
terms. This anharmonic oscillator has been frequently used as a toy model for
the $\Phi^4$-theory \cite{pot}. From a field theoretical point of view,
one of physically interesting quantum states is the Gaussian state, for
it represents a resummation of loop diagrams such as daisies.
Let us consider a model Hamiltonian of the form
\begin{equation}
\hat{H}_{a.h.o.} = \frac{1}{2} \hat{p}^2 + \frac{\omega^2 }{2} \hat{x}^2
+ \frac{\lambda}{4} \hat{x}^4.
\label{anhar osc}
\end{equation}
It was shown in Ref. \cite{bak} that the Hamiltonian
(\ref{anhar osc}) has the following Gaussian state
\begin{equation}
\Psi(x) = \Bigl( \frac{1}{2 \pi \chi^2} \Bigr)^{1/4}
\exp\Bigl[- \Bigl(\frac{1}{4 \chi^4} -
i \frac{\dot{\chi}}{2 \chi} \Bigr)x^2 \Bigr],
\label{gaussian}
\end{equation}
where $\chi$ satisfies the nonlinear equation
\begin{equation}
\ddot{\chi} + \omega^2 \chi + 3 \lambda \chi^3
- \frac{1}{4 \chi^3} = 0.
\label{1-d}
\end{equation}
Eq. (\ref{1-d}) can be easily integrated to yield
\begin{equation}
\frac{1}{2} \dot{\chi}^2 + \frac{\omega^2}{2}
\chi^2 + \frac{3 \lambda}{4} \chi^4 +
\frac{1}{8 \chi^2} = \epsilon,
\label{int eq}
\end{equation}
where $\epsilon$ is a constant of motion. The $\epsilon$ is nothing but
the energy expectation value
\begin{equation}
\langle \Psi \vert \hat{H} \vert \Psi \rangle
= \frac{1}{2} \Bigl[ \dot{\chi}^2 + \omega^2
\chi^2 + \frac{3 \lambda}{2} \chi^4 +
\frac{1}{4 \chi^2} \Bigr] = \epsilon.
\end{equation}

The solution to Eq. (\ref{int eq}) can be found by interpreting
Eq. (\ref{1-d}) as describing a one-dimensional
system of unit mass with an effective potential
\begin{equation}
V_{\rm eff.} = \frac{\omega^2}{2}
\chi^2 + \frac{3 \lambda}{4} \chi^4 +
\frac{1}{8 \chi^2}.
\end{equation}
The effective potential, though one-dimensional, has a centrifugal
term $\frac{1}{8 \chi^2}$ from quantization in addition to the
other classical potential terms.
The energy expectation value has the minimum value at $\chi^2
= \frac{1}{2 \Omega_G}$:
\begin{equation}
\epsilon_{\rm min.} = \frac{\Omega_G}{2} -
\frac{3 \lambda}{16 \Omega_G^2},
\end{equation}
where $\Omega_G$ is determined from the gap equation \cite{pot}
\begin{equation}
\Omega_G^2 = \omega^2 + \frac{3 \lambda}{2 \Omega_G}.
\label{gap eq}
\end{equation}
For an energy above $\epsilon_{\rm min.}$, there are two turning points
$\chi_1$ and $\chi_2$. By setting $y = \chi^2$, we integrate
Eq. (\ref{int eq}) by quadrature method
\begin{equation}
\int_{y_1}^{y} \frac{dy}{\sqrt{ 2 \epsilon y - \omega^2 y^2 -
\frac{3 \lambda}{2} y^3 - \frac{1}{4}}} =
2 t + \varphi_0,
\label{int}
\end{equation}
where $\varphi_0$ is an integration constant. The integrand has two
positive roots $y_2 \geq y_1 > 0$ corresponding to physical turning points
and an unphysical negative root $y_3 < 0$:
\begin{equation}
2 \epsilon y - \omega^2 y^2 -
\frac{3 \lambda}{2} y^3 - \frac{1}{4} =
\frac{3 \lambda}{2} (y_2 - y) (y - y_1) (y - y_3).
\end{equation}
By doing the integral (\ref{int}), we find the solution \cite{gradshteyn}
\begin{equation}
F (\kappa, p) =  \sqrt{\frac{3 \lambda}{2}
(y_2 - y_3)} \Bigl( t + \frac{\varphi_0}{2} \Bigr),
\label{sol}
\end{equation}
where $F(\kappa, p)$ is the first elliptic integral
and
\begin{equation}
\kappa = \arcsin \Bigl\{\sqrt{\frac{(y_2 - y_3)(y - y_1)}{(y_2 - y_1)
(y - y_3)}} \Bigr\},
~~ p = \sqrt{\frac{y_2 - y_1}{y_2 - y_3}}.
\label{para}
\end{equation}
We have thus shown that the Gaussian state (\ref{gaussian}) is
the one-parameter state whose parameter is the energy
expectation value. In order to find the meaning and relation
of this one-parameter Gaussian state (\ref{gaussian}) with the
(approximate) vacuum state, we shall compare it with that of a harmonic
oscillator, which is relatively well understood.

We first consider the case of the harmonic oscillator with $\lambda = 0$
\begin{equation}
\hat{H}_{h. o.} = \frac{1}{2} \hat{p}^2 + \frac{\omega^2 }{2}
\hat{x}^2.
\label{har osc}
\end{equation}
Then one has $p = 0$ in Eq. (\ref{para}) and $F(\kappa, 0) = \kappa$ in
Eq. (\ref{sol}), and the solution (\ref{sol}) reduced to
\begin{equation}
y = \chi^2 = \frac{\epsilon}{\omega^2}
+ \frac{\epsilon}{\omega^2} \sqrt{1 -
\frac{\omega^2}{4 \epsilon^2}} \cos (2 \omega t),
\label{har chi}
\end{equation}
where we have chosen a constant $\varphi_0 =
\frac{\pi}{\omega}$. The connection between the vacuum (minimum energy) state
and the one-parameter Gaussian state (\ref{gaussian}) with $y = \chi^2$ from Eq.
(\ref{har chi}) can be manifestly shown in the Liouville-Neumann approach
\cite{lewis,hartley,kim2,kim1}.
Moreover, the Liouville-Neumann approach provides a better physical
intuition. In this approach one looks for the following
creation and annihilation operators \cite{kim1}
\begin{equation}
\hat{a}^{\dagger} (t) = -i \Bigl(u (t) \hat{p}
- \dot{u} (t) \hat{x} \Bigr), ~\hat{a} (t) = i \Bigl(u^* (t) \hat{p}
-  \dot{u}^* (t) \hat{x} \Bigr),
\label{basis}
\end{equation}
which satisfy the Liouville-Neumann equations
\begin{equation}
i \frac{\partial}{\partial t} \hat{a}^{\dagger} (t)
+ \Bigl[\hat{a}^{\dagger} (t) , \hat{H}_{h.o.} \Bigr] = 0,~
i \frac{\partial}{\partial t} \hat{a} (t)
+ \Bigl[\hat{a} (t) , \hat{H}_{h.o.} \Bigr] = 0,.
\label{ln eq}
\end{equation}
The great advantage of this approach is that the exact
quantum state of either time-independent or time-dependent
harmonic oscillator is an eigenstate of such operators
up to some time-dependent phase factor.
From the fact that any functionals of $\hat{a}(t)$ and
$\hat{a}^{\dagger} (t)$ also satisfy Eq. (\ref{ln eq}),
it follows that the (number) eigenstate of $\hat{N} (t)
= \hat{a}^{\dagger}(t) \hat{a}(t)$ is an exact quantum state
of the harmonic oscillator \cite{lewis}.

The operators (\ref{basis}) satisfy Eq. (\ref{ln eq}) when
$u(t)$ is a complex solution to the classical equation of motion
\begin{equation}
\ddot{u} (t) + \omega^2 u (t) = 0.
\label{har cl eq}
\end{equation}
One can make $\hat{a} (t)$ and $\hat{a} (t)$ the creation
and annihilation operators by imposing the standard commutation
relation $[\hat{a}(t) , \hat{a}^{\dagger}(t)] = 1$,
which is the Wronskian condition
\begin{equation}
\dot{u}^*(t) u(t) - u^* (t) \dot{u} (t) = i.
\label{wron}
\end{equation}
The Gaussian state that is defined by $\hat{a} (t) \vert 0 \rangle = 0$
has the coordinate representation
\begin{equation}
\Psi_u = \Bigl(\frac{1}{2 \pi u^* (t) u (t)} \Bigr)^{1/4}
\exp \Bigl[i \frac{\dot{u}^* (t)}{2 u^* (t)} x^2 \Bigr].
\label{gaussian coord}
\end{equation}
In fact, one can show the equivalence between Eq. (\ref{har cl eq})
and the nonlinear equation (\ref{1-d}) with $\lambda = 0$,
by expressing $u(t)$ in a polar form
$u (t) = \chi (t) e^{- i \theta (t)}$ and by writing Eq. (\ref{wron}) as
$ \dot{\theta} (t) = \frac{1}{2 \chi^2 (t)}$, which is integrated
to be $\theta = \omega t$. It should be noted that a solution with
the particular positive frequency $\omega$,
\begin{equation}
u_0 (t) = \frac{1}{\sqrt{2 \omega}} e^{- i \omega t},
\label{vac sol}
\end{equation}
provides the true vacuum state with the minimum energy.
The vacuum state meets the selection rule of the minimum uncertainty \cite{kim1}.
Eq. (\ref{har cl eq}) being linear and  $u_0^* (t)$ another
independent solution, any superposition of $u_0 (t)$ and $u_0^* (t)$
also constitutes a solution. We may parameterize such coefficients so that
they satisfy  the Wronskian condition (\ref{wron}):
 \begin{equation}
u_{\rm r} (t) = \cosh (r) u_0 (t) + \sinh (r) u_0^* (t).
\label{lin com2}
\end{equation}
By identifying $\chi^2 = u_{\rm r}^* u_{\rm r}$
in Eq. (\ref{har chi}), the squeezing parameter $r$ is found
in terms of the energy as
\begin{equation}
\cosh (r) = \sqrt{\frac{\epsilon}{\omega} +
\frac{1}{2} },~
\sinh (r) = \sqrt{\frac{\epsilon}{\omega} - \frac{1}{2}}.
\end{equation}
The creation and annihilation operators defined in terms of $u_{\rm r} (t)$
\begin{equation}
\hat{a}^{\dagger}_{\rm r} (t) = -i \Bigl( u_{\rm r} (t) \hat{p}
- \dot{u}_{\rm r} (t) \hat{x} \Bigr),~
\hat{a}_{\rm r} (t) = i \Bigl( u^*_{\rm r} (t) \hat{p}
-  \dot{u}^*_{\rm r} (t) \hat{x} \Bigr),
\label{basis2}
\end{equation}
are related with those for the vacuum state defined in terms of $u_0 (t)$
through the Bogoliubov transformation
\begin{eqnarray}
\hat{a}_{\rm r}^{\dagger} (t) &=& - \sinh (r) \hat{a}_0 (t)
+ \cosh (r) \hat{a}^{\dagger}_0 (t),
\nonumber\\
\hat{a}_{\rm r} (t) &=& \cosh (r) \hat{a}_{0} (t)
- \sinh (r) \hat{a}^{\dagger}_{0} (t).
\end{eqnarray}
Hence, the one-parameter Gaussian state for harmonic oscillator
is the squeezed state (squeezed vacuum) of the true vacuum state that
is annihilated by $\hat{a}_0 (t)$ and has $r = 0$ \cite{mandel}.

We now turn to the case of the anharmonic oscillator $(\lambda \neq 0)$.
Years ago Rajagopal and Marshall found a coherent and Gaussian state for
time-independent anharmonic oscillator with
a polynomial potential \cite{rajagopal}, and recently
one (SPK) of the authors used the Liouville-Neumann approach
to find a similar coherent and Gaussian state for time-dependent
anharmonic oscillator \cite{kim9}. The idea of Ref. \cite{kim9} is to find
the following operators
\begin{equation}
\hat{A}^{\dagger} (t) = -i \Bigl(v (t) \hat{p}
- \dot{v} (t) \hat{x} \Bigr), ~\hat{A} (t) = i \Bigl(v^* (t) \hat{p}
-  \dot{v}^* (t) \hat{x} \Bigr),
\label{basis5}
\end{equation}
which satisfy the Liouville-Neumann equations for the full
Hamiltonian (\ref{anhar osc})
\begin{equation}
i \frac{\partial}{\partial t} \hat{A}^{\dagger} (t)
+ \Bigl[\hat{A}^{\dagger} (t) , \hat{H}_{a.h.o.} \Bigr] = 0,~
i \frac{\partial}{\partial t} \hat{A} (t)
+ \Bigl[\hat{A} (t) , \hat{H}_{a.h.o.} \Bigr] = 0.
\label{ln eq5}
\end{equation}
It was found that Eq. (\ref{ln eq5}) is satisfied only when
$v(t)$ is a complex solution to
\begin{equation}
\ddot{v} (t) + \langle 0 (t) \vert \frac{\delta^2}{\delta \hat{x}^2}
V(\hat{x}) \vert 0 (t) \rangle v(t) = 0,
\label{anhar eq}
\end{equation}
where the expectation value is taken with respect to the state
defined by $\hat{A} (t) \vert 0 (t) \rangle = 0$.
For the Hamiltonian (\ref{anhar osc}), Eq. (\ref{anhar eq}) becomes
\begin{equation}
\ddot{v} (t) + \omega^2  v(t) + 3 \lambda
\Bigl(v^* (t) v (t) \Bigr) v (t) = 0.
\label{cl eq}
\end{equation}
It is worthy of noting that the same mean-field equation
(\ref{cl eq}) can also be obtained by expressing the Hamiltonian
(\ref{anhar osc}) in terms of the operators in Eq. (\ref{basis5}),
by arraying it in a normal ordering, and by truncating it up to the quadratic terms
\begin{eqnarray}
\hat{H}_G &=& \Bigl[\dot{v}^* \dot{v} + \omega^2
v^* v + 3 \lambda (v^* v)^2 \Bigr] \Bigl(\hat{A}^{\dagger} \hat{A} +
\frac{1}{2} \Bigr) - \frac{3 \lambda}{4} (v^* v)^2
\nonumber\\
&&+ \frac{1}{2} \Bigl[\dot{v}^{*2} \dot{v} + \omega^2
v^{*2} + 3 \lambda (v^* v) v^{*2} \Bigr] \hat{A}^{\dagger 2}
+ \frac{1}{2} \Bigl[\dot{v}^{2} \dot{v} + \omega^2
u^{2} + 3 \lambda (v^* v) v^{2} \Bigr] \hat{A}^{2},
\end{eqnarray}
and finally by solving Eq. (\ref{ln eq}) with
the approximate Hamiltonian $\hat{H}_G$.
A particular solution to Eq. (\ref{cl eq}) with $\Omega_G$
determined by the gap equation (\ref{gap eq}),
\begin{equation}
v_0 (t) = \frac{1}{\sqrt{2 \Omega_G}} e^{- i \Omega_G t},
\label{pos-neg}
\end{equation}
provides an approximate vacuum (minimum energy) state.
As for the harmonic oscillator case, by setting $v (t) =
\chi (t) e^{ - i \theta (t)}$ and using $\dot{\theta} (t) =
\frac{1}{2 \chi^2 (t)}$ from the commutation relation
$[\hat{A} (t), \hat{A}^{\dagger} (t)] = 1$, one can show
that Eq. (\ref{cl eq}) becomes identically the nonlinear equation
(\ref{1-d}) with $\lambda \neq 0$.

As a special case, we consider the weak coupling constant
$(\lambda \ll \omega^2)$, and from now on compute all quantities to
the linear order of $\lambda$. From the three roots approximately given by
\begin{eqnarray}
y_1 &=& \frac{\epsilon}{\omega^2} \Bigl[1 - \sqrt{1 -
\frac{\omega^2}{4 \epsilon^2}} \Bigr]
- \frac{3 \lambda}{16 \omega^6} \Bigl[
16 \epsilon^2 - \omega^2 - \frac{16 \epsilon^2 - 3 \omega^2
}{\sqrt{1 - \frac{\omega^2}{4 \epsilon^2}}}\Bigr]
+ {\cal O} (\lambda^2),
\nonumber\\
y_2 &=& \frac{\epsilon}{\omega^2} \Bigl[1 + \sqrt{1 -
\frac{\omega^2}{4 \epsilon^2}} \Bigr]
- \frac{3 \lambda}{16 \omega^6} \Bigl[
16 \epsilon^2 - \omega^2 + \frac{16 \epsilon^2 - 3 \omega^2
}{\sqrt{1 - \frac{\omega^2}{4 \epsilon^2}}}\Bigr]
+ {\cal O} (\lambda^2),
\nonumber\\
y_3 &=& - \frac{2 \omega^2}{3 \lambda} -
\frac{\epsilon}{\omega^2} +
\frac{6 \epsilon^2 \lambda}{\omega^6} \Bigl[
1 - \frac{\omega^2}{16 \epsilon^2} \Bigr]
+ {\cal O} (\lambda^2),
\end{eqnarray}
the solution (\ref{sol}) is found to be
\begin{eqnarray}
y = \chi^2 &=& \Bigl[\frac{\epsilon}{\omega^2}
- \frac{9 \lambda}{16 \omega^6} \bigl(8 \epsilon^2
-  \omega^2 \bigr) + {\cal O} (\lambda^2) \Bigr]
\nonumber\\
&+& \Bigl[\frac{\epsilon}{\omega^2} \sqrt{1 -
\frac{\omega^2}{4 \epsilon^2}}
- \frac{3 \lambda}{16 \omega^6}
\frac{16 \epsilon^2 - 3 \omega^2
}{\sqrt{1 - \frac{\omega^2}{4 \epsilon^2}}}
+ {\cal O} (\lambda^2)\Bigr] \cos (2 \Omega t),
\end{eqnarray}
where
\begin{equation}
\Omega = \omega + \frac{3 \lambda}{4 \omega}
\Bigl[\frac{2 \epsilon}{\omega^2} + \sqrt{1
- \frac{\omega^2}{4 \epsilon^2}} \Bigr]
+ {\cal O} (\lambda^2).
\label{new freq}
\end{equation}
It should be remarked that near the minimum energy
$\epsilon_{\rm min.}$, $\Omega$ is equal to $\Omega_G$
from the gap equation (\ref{gap eq}):
\begin{equation}
\Omega = \omega + \frac{3 \lambda}{4 \omega^2} + {\cal O} (\lambda^2)
 = \Omega_G.
\end{equation}
This can be understood intuitively because near the minimum
of potential the anharmonic oscillator is approximated by a
harmonic potential with corrections from the quartic term.
We are thus able to show the following linear superposition
\begin{equation}
v_{\rm r} (t) = \cosh (r) v_0 (t) + \sinh (r) v_0^* (t),
\label{lin com}
\end{equation}
where
\begin{eqnarray}
\cosh (r) &=& \sqrt{\frac{\Omega y_2}{2} +
\frac{\Omega (y_2 - y_1 )^2}{8 y_3}} +
\sqrt{\frac{\Omega y_1}{2} +
\frac{\Omega (y_2 - y_1 )^2}{8 y_3}},
\nonumber\\
\sinh (r) &=& \sqrt{\frac{\Omega y_2}{2} +
\frac{\Omega (y_2 - y_1 )^2}{8 y_3}} -
\sqrt{\frac{\Omega y_1}{2} +
\frac{\Omega (y_2 - y_1 )^2}{8 y_3}}.
\end{eqnarray}
Hence, to the linear order of the coupling constant (weak coupling
limit),
the Bogoliubov transformation  between $\hat{A}_{\rm r} (t),
\hat{A}^{\dagger}_{\rm r} (t)$ and $\hat{A}_0 (t), \hat{A}^{\dagger}_0 (t)$
holds
\begin{eqnarray}
\hat{A}_{\rm r}^{\dagger} (t) &=& - \sinh (r) \hat{A}_0 (t)
+ \cosh (r) \hat{A}^{\dagger}_0 (t),
\nonumber\\
\hat{A}_{\rm r} (t) &=& \cosh (r) \hat{A}_{0} (t)
- \sinh (r) \hat{A}^{\dagger}_{0} (t),
\end{eqnarray}
where
\begin{equation}
\hat{A}^{\dagger}_{\rm r} (t) = -i \Bigl( v_{\rm r} (t) \hat{p}
- \dot{v}_{\rm r} (t) \hat{x} \Bigr),~
\hat{A}_{\rm r} (t) = i \Bigl( v^*_{\rm r} (t) \hat{p}
-  \dot{v}^*_{\rm r} (t) \hat{x} \Bigr).
\label{basis3}
\end{equation}
Therefore, one-parameter Gaussian state for the weak
coupling constant is the squeezed vacuum of the Gaussian state
with $\Omega_G$.

In summary, we have found a one-parameter (energy expectation value)
dependent Gaussian state for a quartic anharmonic oscillator.
This Gaussian state is not in general a squeezed vacuum of the vacuum state
through Bogoliubov transformation. This fact is in strong contrast with
the harmonic oscillator case where any Gaussian state is always a squeezed state
of the true vacuum. Only for the weak coupling constant,
this Gaussian state is the squeezed state of the
vacuum state. However, for the strong coupling constant, one can not
obtain the Gaussian state through a linear Bogoliubov transformation
of the vacuum state.
The one-parameter Gaussian state with an energy much higher than
the vacuum energy represents a different kind of condensation of bosonic particles
in a nonlinear way. Though not shown explicitly in this paper,
the operator formalism for the nonlinear transformation
between the vacuum state and the other Gaussian states
would be interesting and and will be addressed in a future publication.

\acknowledgements

We would like to thank Dr. Dongsu Bak, Prof. Sung Ku
Kim, Prof. Kwang-Sup Soh, and Prof. Jae Hyung Yee for many useful
discussions and comments. This work was supported by the
Non-Directed Research Fund, Korea Research Foundation, 1997.

\end{document}